# Impact of Green Marketing Strategy on Brand Awareness: Business, Management, and Human Resources Aspects

**Mahdi Nohekhan**

Executive Management Group, Islamic Azad University, Shahrood Branch,

**Mohammadmahdi Barzegar**

Sheldon B. Lubar College of Business, University of Wisconsin Milwaukee

barzega3@uwm.edu

**Abstract**

Given the move towards industrialization in societies, the increase in dynamism and competition among companies to capture market share, raising concerns about the environment, government, and international regulations and obligations, increased consumer awareness, pressure from nature-loving groups, etc., organizations have become more attentive to issues related to environmental management. Over time, concepts such as green marketing have permeated marketing literature, making environmental considerations one of the most important activities of companies. To this end, this research examines the impact of green marketing strategy on brand awareness (case study: food exporting companies). The population of this research consists of 345 employees and managers of companies like Kalleh, Solico, Pemina, Sorbon, Mac, Pol, and Castle, from which 182 individuals were randomly selected as the sample using Cochran's formula. This research is practical, and the required data have been collected through a survey and a questionnaire. The research results indicate that (1) green marketing strategy significantly affects brand awareness. (2) Green products have a significant positive effect on brand awareness. (3) Green promotions have a significant positive effect on brand awareness. (4) Green distribution has a significant positive effect on brand awareness. (5) Green pricing has a significant positive effect on brand awareness.

**Introduction**

Practical attention to ethical, social, and environmental issues in business dates back to the 1960s. However, interest in such matters has grown increasingly over the past 20 years. The evolutionary path in this research area has witnessed the integration of various theories. Although research with a sustainability background is detailed and diverse, most studies to date have presented one or more than five key items. First is the diversity of sustainability, which, along with internal and external factors, represents the adaptation of companies to environmentally or socially friendly strategies. The second issue is sustainability management, which focuses on a company's sustainability performance and strategies. The third issue is that the results of sustainability performance focus on financial, market, customer, operational, shareholder, and community aspects. Despite this, recent studies in this area state that the evidence on the results of corporate sustainability efforts has remained inconclusive. Green marketing strategy includes all activities designed to create and facilitate exchanges to satisfy human needs and wants so that fulfilling these needs and wants involves minimal harmful and destructive environmental impacts. Therefore, given the challenge facing most companies, green marketing strategies are implemented through developing and creating green marketing programs (Karsten, Iris, 2014; Sepahvand et al., 2023; Nawaser et al., 2023).

Sharma and Vredenburg (1998) concluded that an active environmental strategy might lead to developing unique and valuable organizational capabilities for competition. In a study by Kabiraj and colleagues (2010), they referred to the business community's changing economic and social awareness and, consequently, the importance of creating an environmentally compatible business model within organizations. Similarly, Olsen (2008) stated that business managers and corporate decision-makers often have not adopted significant green opportunities in the strategic field of the organization. Sharma (2000) examined the relationship between managerial interpretations of environmental issues and a company's environmental strategy selection. In his study, he highlighted the importance of organizational and managerial factors that influence the choice of strategy (Moeini and colleagues, 2014; Nawaser et al., 2023).

The issue of environmental preservation has prompted consumers to reassess their purchasing decisions, leading to a growing preference for products that adhere to environmental standards. This shift has spurred the rise of green marketing, an organization's strategic initiative to provide eco-friendly products. Such an approach appeals to environmentally conscious consumers and opens avenues for sustainable competitive advantage, as highlighted by Naduviya and colleagues in 2015.

In this evolving market landscape, employee innovation emerges as a pivotal factor, especially in the context of green marketing. Daneshmandi et al. (2023) emphasize the critical role of innovation within organizations, underscoring the need to explore factors that foster individual creativity and team collaboration in this domain. This is where the concept of job satisfaction gains prominence. A satisfied workforce, particularly within marketing departments, can significantly influence the success of green marketing initiatives.

In recent years, the intricate dynamics between job satisfaction and its impact on innovation have garnered attention in organizational studies. For instance, Bai et al. (2023) conducted a study focusing on female employees, revealing a positive correlation between job satisfaction and enhanced performance and innovation. This finding is particularly relevant in green marketing, where employee ingenuity is indispensable. Without considering and nurturing individual job satisfaction levels, advancing green marketing strategies could pose a challenge for businesses. This interplay between employee satisfaction, innovation, and green marketing strategies forms a crucial triad in achieving marketing success in today's environmentally conscious business environment.

In sustainability literature, green marketing refers to the performances, policies, and marketing programs that specifically address concerns about the natural environment, generate revenue, and provide results that meet individual and organizational objectives for a product or production line (Konstantin and colleagues, 2013). Therefore, green marketing programs are designed to minimize negative impacts on the natural environment.

Green marketing was first introduced in the late 1970s by the American Marketing Association (AMA). The first workshop on environmental marketing was held in 1975, leading to the publication of the first book on the subject, "Ecological Marketing" by Kenner Hinton in 1976. Germany was the first Eco member country to label environmentally friendly products in 1978. Green marketing became widely discussed in the late 1980s and early 1990s. Since then, the definition of green marketing has been refined and divided into three main parts as follows:

- Marketing of products with the creation of a safe environmental market.
- Development of marketing products designed to minimize negative impacts on the physical environment or improve quality.
- Organizational efforts to produce, promote, package, and modify products in a way that responds to environmental concerns (Mohajan, 2012).

Table 1: Evolution of Environmental Interests.

| Factor | Environmentalism in the 1970s | Green Marketing in the Present Era |
|---|---|---|
| **Emphasis** | On environmental issues | **On understanding issues related to social, economic, technical, and legal systems** |
| **Geographical Focus** | On local issues (like air pollution) | **On global issues (like global warming)** |
| **Identity** | Closely associated with other anti-corporation debates | **A distinct movement from anti-corporation debates** |
| **Source of Support** | An elite, educated stratum, and others on the fringes of society | **A broad social base** |
| **Basis for Comprehensive Promotion** | Use of forecasts related to growth hazards on environmental issues | **Use of evidence based on the current environment (such as the ozone layer gap)** |
| **Attitude Towards Businesses** | The problem is from the businesses themselves. Businesses are generally seen as enemies of the environment. | **Businesses are seen as part of the solution and should engage more.** |

Branding is a facet of marketing that emerged in the nineteenth century with the advent of packaged products. Industrialization transformed the production of local soap into centralized soap manufacturing factories. When products were shipped, factories had to mark their emblems on the cargo to identify which producer the products belonged to. Mass production ensued, and such production volume necessitated access to a broader market.

Manufacturers soon realized that simply packaging a soap bar was not enough to compete with local products and that a different packaging was needed. Packaged goods are needed to convince the market that these products could meet consumer needs better than locally produced goods. Companies began to name their products to differentiate between local products and mass-produced, non-local products, thus creating the first brand names. Campbell's soap, Coca-Cola, and Quaker Oats were among the first commercially named products to familiarize consumers with the product.

The American Marketing Association (1960) defines a brand as "a name, term, sign, symbol, or design, or a combination of them, intended to identify the goods and services of one seller or group of sellers and to differentiate them from those of competitors." A brand is a product or service that adds dimensions that distinguish it from others. These distinctions can be functional, tangible, or even intangible. Philip Kotler also provides a similar definition to what the American Marketing Association has proposed. A brand is a name, phrase, term, symbol, design, or a combination thereof whose purpose is to introduce a product or service offered by a seller or a group of sellers and thereby distinguish it from the products of competing companies (Kotler, 2012).

Brand awareness is the potential buyer's ability to recognize and recall a brand as a member of a specific product category. In other words, a product category (like cars) evokes a specific brand, such as Mercedes. Consumer-based brand equity is created when the consumer has a high level of awareness and familiarity with the brand and holds unique, favorable, and strong associations in mind (Kotler, 2012).

(1) A basis for other associations that may be of interest, (2) establishment of a familiar link, (3) a sign of enduring commitment or attention. Keller (2003) stated that brand familiarity plays a significant role in consumer purchase decisions and carries the advantage of learning, attention, and choice. A brand traditionally consumed by a family over the years creates a high level of awareness in the consumer's mind, a form of learning in the home environment due to the long-term consumption of brands within the family. This awareness could significantly strengthen the consumer's learning in identifying a brand within its product class to such an extent that those brands might also appear on the shopping lists of future generations. According to McDonald and Sharp (2003), brand awareness is one of the main parts of the famous advertising effect hierarchy models and one of the important goals of communicative activities for marketing managers, who use this concept to measure the effectiveness of marketing and advertising activities. When a customer is familiar with a large number of brands that align with their criteria, it seems unlikely that they will make much effort to search for information about unfamiliar competitor brands, and one of the important functions is to increase the perceived quality of the product or service by the consumer.

- Researchers such as Ramezaniyan, Esmaeilpour, and Tondkar (2010) investigated the impact of the green marketing mix on the consumer decision-making process for low-consumption light bulbs. While each study has very valuable results, they also have limitations. Another necessity is that the majority of studies conducted in this area have been related to Western countries, focusing on consumers of those countries, while consumers in other parts of the world may have different attitudes and beliefs towards marketing activities; therefore, the results presented in these studies may not be valid for Iranian consumers who have different cultures, beliefs, and attitudes compared to Western consumers (Ranaei et al., 2012).

- Avan (2011) investigated green marketing: marketing strategies of Swedish energy companies. Commercial companies have become aware of the need to preserve the environment and have incorporated marketing strategies for environmental protection as a social responsibility into their agendas. In this paper, the reasons companies have adopted green marketing (related to producing environmentally friendly products) were analyzed, and findings were obtained. This article is based more on relevant writings and reviews of empirical studies, aiming to understand the importance of integrating environmentally based marketing within companies. It also states that environmental labels on goods distinguish environmentally friendly producers from ordinary ones. The key finding of this article is that large companies cannot be completely detached from green business and must equally participate in social programs. These companies have been empowered to decide to what extent they are willing to incorporate green marketing strategies into their overall company programs. This environmental pattern has been adapted from the study of existing business literature.

- Chan (2015) investigated the factors affecting the green purchasing behavior of Chinese consumers. This study examined the cultural and psychological factors affecting the purchasing behavior of Chinese consumers towards green products. Hypotheses were tested using a conceptual model and through surveys. Results showed that individuals' perception of green products and their purchase history affect their consumption levels.

**Research Design**

Research Methodology and Objective: This study is based on a practical objective and is a descriptive survey method of the correlational type.

Sampling Method and Data Collection: The statistical population of the current research includes managers and employees of food export companies such as Kalleh, Solico, Pemina, Sorbon, Mac, Pol, and Castle, who were working during the years 1394-1395 SH. According to the obtained statistics, their total number is 345 individuals, of which, based on Cochran's formula, about 182 individuals are calculated for the sample size.

Statistical Methods and Data Analysis Technique: In this research, the green marketing strategy questionnaire derived from Imam Gholi's study (2014) (which includes 7 questions about green product, 8 questions about green promotion, 3 about green distribution, 5 about green price) and Aaker's brand awareness model questionnaire (1991) (which includes 8 questions) were used. Since this research studies a sample of the statistical population and the data of this research are based on a five-point Likert scale, ranging from "strongly disagree" to "strongly agree", the scoring of the questions is also calculated from 1 to 5. After collecting, the data obtained from the questionnaires were transferred to SPSS software version 19 raw data sheets and analyzed using descriptive statistics (frequency, percentage, tables) and inferential statistics (Pearson correlation).

Reliability and Validity of the Research Questionnaires: To examine the validity of the measurement tools used in this research, the opinions of university professors and experts were used. Also, since the items considered in this questionnaire are based on standard questionnaires, the questionnaire is deemed to have satisfactory validity. In the present study, a preliminary sample including 30 questionnaires was pre-tested, and then the reliability coefficient was calculated using Cronbach's alpha method with the data obtained from these questionnaires and the statistical software SPSS, which showed that the Cronbach's alpha coefficient for the green marketing strategy questionnaire is 0.972, and for the components of the green marketing strategy questionnaire (green product 0.945, green promotion 0.881, green distribution 0.806, green price 0.912), and for the brand awareness questionnaire (0.911). Therefore, it can be said that the questionnaires have high reliability, correlation, and internal credibility.

Conceptual Model of Research and Hypotheses: In this research, following the models of Hick and Yalden (2013) and Moskowitz and colleagues (2015), hypotheses are tested. The conceptual model of the current research is researcher-constructed and results from the review of various models and literature (Figure 1).

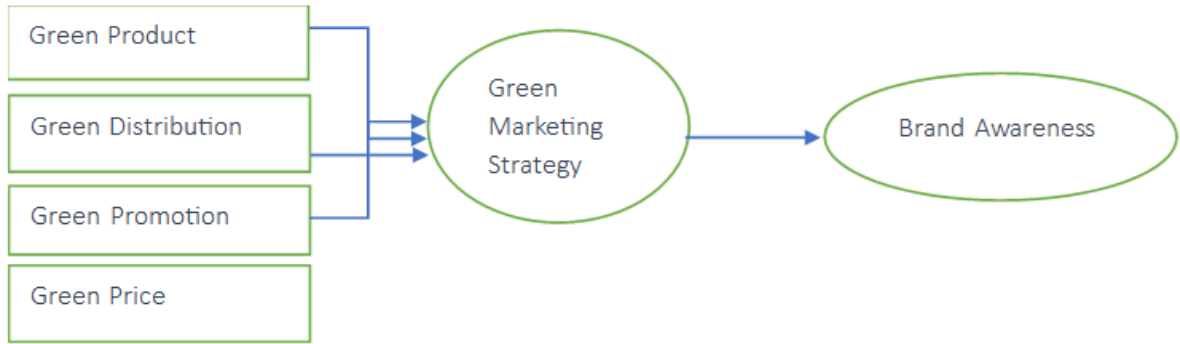

Main Hypothesis of the Research: Green marketing strategy significantly affects brand awareness.

Sub-hypothesis 1: Green product has a significant effect on brand awareness.

Sub-hypothesis 2: Green promotion has a significant effect on brand awareness.

Sub-hypothesis 3: Green distribution has a significant effect on brand awareness.

Sub-hypothesis 4: Green pricing has a significant effect on brand awareness.

**Findings**

The frequency and percentage frequency of the research sample group based on gender, educational level, and service history are presented in Table (2).

Table 2: Frequency and Percentage Frequency of the Research Sample Group.

| Variable | Subgroups | Frequency | Percentage Frequency |
|---|---|---|---|
| **Gender** | Male | 103 | 56.60% |
| | Female | 79 | 43.40% |
| **Age** | Less than 30 | 53 | 29.10% |
| | 31 to 40 | 61 | 33.50% |
| | 41 to 50 | 46 | 25.30% |
| | More than 50 | 22 | 12.10% |
| **Education** | Diploma and below | 31 | 17.00% |
| | Associate degree | 43 | 23.60% |
| | Bachelor's degree | 59 | 32.40% |
| | Graduate degree | 49 | 26.90% |
| **Work Experience** | Less than 1 year | 21 | 11.50% |

| | 1 to 5 years | 34 | 18.70% |
| --- | --- | --- | --- |
| | 6 to 10 years | 78 | 42.90% |
| | More than 10 years | 49 | 26.90% |

As shown in Table 2, the majority of the study sample in terms of gender is male; in terms of educational level, it is those with a bachelor's degree. In terms of work experience, it is those with 6 to 10 years of experience, and in terms of age, it is those aged between 31 to 40 years.

Table 3: Pearson Correlation Test (Main Hypothesis)

| | | Green Marketing Strategy | Brand Awareness |
| --- | --- | --- | --- |
| **Green Marketing Strategy** | Pearson Correlation Coefficient | 1 | .760 |
| | Significance Level | | .000 |
| | Sample Size | 182 | 182 |
| **Brand Awareness** | Pearson Correlation Coefficient | .760 | 1 |
| | Significance Level | .000 | |
| | Sample Size | 182 | 182 |

The results of the Pearson correlation test presented in Table 3 indicate that the significance level is equal to zero and less than 0.05, which signifies a meaningful relationship between the green marketing strategy and brand awareness. Therefore, the main hypothesis of the research is confirmed. Additionally, the Pearson correlation coefficient is 0.760, indicating a positive and significant effect of the green marketing strategy on brand awareness.

Table 4: Pearson Correlation Test (First Sub-Hypothesis)

| | | Brand Awareness | Green Product |
| --- | --- | --- | --- |
| **Brand Awareness** | Pearson Correlation Coefficient | 1 | .800 |
| | Significance Level | | .000 |
| | Sample Size | 182 | 182 |
| **Green Product** | Pearson Correlation Coefficient | .800 | 1 |
| | Significance Level | .000 | |
| | Sample Size | 182 | 182 |

The results of the Pearson correlation test presented in Table 4 indicate that the significance level is equal to zero and less than 0.05, which signifies a meaningful relationship between green product and brand awareness. Therefore, the first sub-hypothesis of the research is confirmed. Additionally, the Pearson correlation coefficient is 0.800, indicating a positive and significant effect of the green product on brand awareness.

Table 5: Pearson Correlation Test (Second Sub-Hypothesis)

|  |  | Brand Awareness | Green Promotion |
|---|---|---|---|
| **Brand Awareness** | Pearson Correlation Coefficient | 1 | .910 |
|  | Significance Level |  | .000 |
|  | Sample Size | 182 | 182 |
| **Green Promotion** | Pearson Correlation Coefficient | .912 | 1 |
|  | Significance Level | .000 |  |
|  | Sample Size | 182 | 182 |

The results from the Pearson correlation test shown in Table 5 indicate that the significance level is equal to zero and less than 0.05, demonstrating a meaningful relationship between green promotion and brand awareness. Consequently, the second sub-hypothesis of the research is confirmed. Moreover, the Pearson correlation coefficient is 0.910, indicating a positive and significant effect of green promotion on brand awareness.

Table 6: Pearson Correlation Test (Third Sub-Hypothesis)

|  |  | Brand Awareness | Green Distribution |
|---|---|---|---|
| **Brand Awareness** | Pearson Correlation Coefficient | 1 | .730 |
|  | Significance Level |  | .000 |
|  | Sample Size | 182 | 182 |
| **Green Distribution** | Pearson Correlation Coefficient | .730 | 1 |
|  | Significance Level | .000 |  |
|  | Sample Size | 182 | 182 |

The results from the Pearson correlation test presented in Table 6 indicate that the significance level is equal to zero and less than 0.05, which signifies a meaningful relationship between green distribution and brand awareness. Therefore, the third sub-hypothesis of the research is confirmed. The Pearson correlation coefficient is also 0.730, indicating a positive and significant effect of green distribution on brand awareness.

Table 7: Pearson Correlation Test (Fourth Sub-Hypothesis)

|  |  | Brand Awareness | Green Distribution |
|---|---|---|---|
| **Brand Awareness** | Pearson Correlation Coefficient | 1 | .847 |
|  | Significance Level |  | .000 |
|  | Sample Size | 182 | 182 |
| **Green Pricing** | Pearson Correlation Coefficient | .847 | 1 |
|  | Significance Level | .000 |  |

| | Sample Size | 182 | 182 |
|---|---|---|---|

The results from the Pearson correlation test shown in Table 7 indicate that the significance level is equal to zero and less than 0.05, demonstrating a meaningful relationship between green pricing and brand awareness. Consequently, the fourth sub-hypothesis of the research is confirmed. Furthermore, the Pearson correlation coefficient is 0.847, indicating a positive and significant effect of green pricing on brand awareness.

Table 8: Research Hypotheses Examination Results

| Hypothesis | Pearson Coefficient | Confirmation/Rejection | Type of Relationship |
|---|---|---|---|
| Main | 0.76 | Confirmed | Positive |
| Sub 1 | 0.8 | Confirmed | Positive |
| Sub 2 | 0.91 | Confirmed | Positive |
| Sub 3 | 0.73 | Confirmed | Positive |
| Sub 4 | 0.847 | Confirmed | Positive |

**Discussion and Conclusion**

This research aims to analyze the impact of green marketing strategies on brand awareness in Tehran's food export companies. The results of hypothesis testing revealed that green marketing strategy, green products, green promotion, green distribution, and green pricing positively influence brand awareness. As indicated by Seyed Yahya et al. (2012), their research showed that among the four variables – attitude towards purchasing green products, normative beliefs, perceived behavioral control, and past experiences with green products – there is a positive relationship with the inclination to buy these products and their contribution to the purchase attitudes of individuals. Moreover, the findings of this hypothesis align with the research of Hick and Yalden (2013), Moghadam and Amirhosseini (2015), and Soori et al. (2015).

In the company's marketing programs, various methods of introducing green products, such as free samples and product sampling, are recommended to physically engage customers (through their five senses) with the product features and its distinction from other products.

Additionally, it is suggested that the methods used by the company for environmental protection and support in its activities be displayed using tangible models, mockups, or through photos, brochures, and advertising posters to articulate the benefits of using environmentally friendly products. These should be understandable and well-explained for customers, considering the cultural level of different social strata and consumers.

Creating, participating in, and supporting environmental organizations and active NGOs in this field can also significantly contribute to brand awareness, forging a link between industry and the environment and promoting green industry.